\documentstyle{article}
\newcommand{\be}{\begin{equation}}
\newcommand{\ee}{\end{equation}}
\begin{document}

\title{Marshall-Peierls sign rule for excited states of the frustrated
$J_1$--$J_2$ Heisenberg antiferromagnet}

\author{Andreas Voigt and Johannes Richter \\
Institut f\"ur Theoretische Physik\\
Otto-von-Guericke-Universit\"at Magdeburg\\
Postfach 4120, 39106 Magdeburg, Germany\\
Telefon: +49-391-6712473\\
Fax: +49-391-6711131 \\
E-mail: Andreas.Voigt@Physik.Uni-Magdeburg.de 
\and
Nedko B. Ivanov \\
Georgi Nadjakov Institute for Solid State Physics\\
Bulgarian Academy of Sciences\\
72 Tzarigradsko chaussee blvd.\\
1784 Sofia, Bulgaria}

\date{\today}
\maketitle

\begin{abstract}
We present analytical and numerical calculations for
some exited states of the frustrated $J_1$--$J_2$ spin-${1\over2}$
Heisenberg model for linear chains and square lattices. We consider the
lowest eigenstates in the subspaces determined by the eigenvalue $M$ of
the spin operator $S_{total}^z$. Because of the reduced number of Ising
basis states in the subspaces with higher $M$ we are able to diagonalize
systems with up to $N=144$ spins. We find evidence that the
Marshall-Peierls sign rule survives for a relatively large frustration
parameter $J_2$.
\end{abstract}

\section{Introduction}
We study the Marshall-Peierls sign rule (MPSR) for the frustrated
spin-${1\over2}$ Heisenberg antiferromagnet
\be
\label{H12}
{\rm \hat H} =J_1 \sum_{\langle {\bf NN} \rangle} {\bf S_i} {\bf S}_{\bf
j} + J_2 \sum_{\langle {\bf NNN} \rangle} {\bf S_i} {\bf S}_{\bf j},
\ee
where $\langle {\bf NN}\rangle$ and $\langle {\bf NNN}\rangle$ denote
nearest-neighbor and next-nearest-neighbor bonds on a linear chain or a
square lattice.

According to Marshall's early work in 1955 \cite{Marshall:1955} we know
the relative phases of the Ising basis states building the ground state
wave function of a bipartite spin-${1\over2}$ Heisenberg antiferromagnet
(MPSR, for more details see below). Later on Lieb and Mattis generalized
the theorem to arbitrary site spins and bipartite lattices without
translational symmetries \cite{Lieb:1961,Lieb:1962}. The knowledge of
the sign is of great importance for the construction of variational wave
functions (see e.g. \cite{Richter:zpb93}) and for quantum Monte Carlo
procedures which may have the so-called minus-sign problem if the MPSR
is violated (see e.g.\cite{Raedt}). In particular, the possible
violation of the MPSR in frustrated systems is a serious problem for
variational and quantum  Monte Carlo procedures.

For the considered model (\ref{H12}) the MPSR can be proved for $J_2 \le
0$. In the frustrated model ($J_2 > 0$) the MPSR can be destroyed
\cite{Richter:zpb93,Kitatani:92,Richter:epl94}. However, in a recent
work \cite{Richter:zpb93,Richter:epl94} we have presented general
arguments that the MPSR may survive for relatively large $J_2$. These
arguments are based on exact diagonalization results for small clusters
(number of sites $N \le 24$), as well as on the spin wave approximation.
Following these arguments Zeng and Parkinson \cite{Zeng:prb95} use the
MPSR as a new way of investigating the spatial periodicity in the ground
state of frustrated spin chains. Furthermore, they studied the breakdown
of the MPSR as a function of the chain length and the frustrating $J_2$.
By finite size extrapolation they estimated a finite critical value for
$J_2$ for an infinite chain of about $0.03J_1$ where the MPSR is
violated in the ground state.

Because of the exponential growth of the number of basis states the
direct numerical calculation of the singlet ground state is limited to
small clusters and the conclusions obtained from small systems seem not
to be quite reliable.

In this paper we exploit the observation that the MPSR holds not only
for the singlet ground state but also for every lowest eigenstate in any
subspace with higher quantum number $M \le {N\over2}$ of the z-component
of the total spin. In these subspaces the number of basis states is
reduced and it is possible to diagonalize much larger systems. With this
data an approximation to the thermodynamic limit is more reliable. Below
we present data up to $N\le144$ and we can conclude that the MPSR
survives indeed a finite frustration $J_2$ at least for states with
higher quantum numbers $M$.

\section{Marshall Peierls sign rule}
In the unfrustrated limit $J_2=0$ the lowest eigenstate of the
Hamiltonian (\ref{H12}) in each subspace of fixed eigenvalue $M$ of the
spin operator $S_{total}^z$ reads
\be
\label{GZ:Phi}
\Psi_{M} = \sum_{m}{c_{m}^{(M)}|m \rangle} \hspace{0.2cm},
\hspace{0.5cm} c_{m}^{(M)}>0 \hspace{0.2cm}.
\ee
Here the Ising states $|m\rangle$ are defined by
\be
\label{m}
|m\rangle \equiv (-1)^{S_A-M_A}|m_1\rangle \otimes |m_2 \rangle \otimes
\cdots \otimes |m_N\rangle \hspace{0.2cm},
\ee
where $|m_i\rangle,\hspace{0.2cm}i=1,\cdots,N$, are the eigenstates of
the site spin operator $S_{i}^{z}$ ($ -s_{i} \leq m_{i} \leq s_i$),
$S_A= \sum_{i\in A}s_i$, $M_{A(B)}=\sum_{i\in A(B)}m_i$, $M=M_A+M_B$.
The lattice consists of two equivalent sublattices $A$ and $B$.
$s_i\equiv s$, $i=1,\cdots,N$, are the site spins. The summations in
Eq.(\ref{GZ:Phi}) are restricted by the condition $\sum_{i=1}^N m_i=M$.
The wave function (\ref{GZ:Phi}) is not only an eigenstate of the
unfrustrated Hamiltonian ($J_2=0$) and $S_{total}^z$ but simultaneously
of the square of the total spin ${\bf S}_{total}^2$ with quantum number
$S=\mid \!M \! \mid$. Because $c_m^{(M)}>0$ is valid for each $m$ from
the basis set (\ref{m}) it is impossible to build up other orthonormal
states without using negative amplitudes $c_m^{(M)}$. Hence the ground
state wave function $\Psi_M$ is nondegenerated.

For the lowest energy eigenvalues $E(S)$ belonging to the subspace $M$
we have the Lieb-Mattis level-ordering
\be
\label{e}
E(S)<E(S+1) \hspace{0.2cm}, \hspace{0.2cm} S\geq 0 \hspace{0.5cm},
\ee
i.e. the ground state is a singlet. Note that this level ordering might
be violated by frustration. However, a lot of numerical calculations
show the same level ordering also for strongly frustrated systems
\cite{Bernu92,Richter95}.

\section{Analytic Solutions}
Now we include frustration ($J_2>0$). In the subspace with maximum
$M=N/2$ the MPSR is never violated in any dimension. Here the only
possible state is the fully polarized ferromagnetic state which does not
change with increasing $J_2$.

In the next subspace $M=(N/2)-1$ analytic solutions are found for linear
chains and square lattices. In this subspace we deal with the so-called
one-magnon state, where the wave function can be expressed as a
Bloch-wave with a given $\vec k$.

{\it Linear Chain} - In this case the solution can be found by comparing
the energies as a function of the wave vector $\vec k$.
\be
E(k) = J_1 ( {N\over 4} -1 + \cos(k) ) + J_2 ( {N\over 4} -1 +  \cos(2k) )
\ee
with $\vec k = {2 \pi\over N} i , i=0,\pm1,\pm2,\ldots,+{N\over2} $.
The comparison of $E(\pi)$ and $E({2 \pi\over N}({N\over2}-1))$ yields the
equation for the critical $J_2$
\be
J_2^c = J_1 {1 + \cos \left[ \pi (1- {2 \over N}) \right] \over
                1 - \cos \left[ 2\pi (1- {2 \over N}) \right] } .
\ee
In the limit $N \rightarrow \infty$ one obtains $J_2^c={1\over4}J_1$.

{\it Square lattice} -
For small $J_2$ in the considered subspace  the lowest energy is
obtained for  $\vec k=(\pi,\pi)$ and reads $E_1=J_1(N-8)+J_2N$. The
corresponding eigenfunction fulfills the MPSR. For larger $J_2$ we have
to distinguish between two cases: (a) If the number of spins in the
sublattice $N/2$ is even we find a transition at $J_2=(J_1/2)$ to a
twofold degenerated ground state with $\vec k =(\pi,0)$ or $\vec k
=(0,\pi)$ with an energy $E_2=J_1(N-4)+J_2(N-8)$. This state violates
the MPSR, i.e. we have $J_2^c={1\over2}J_1$. Notice, that  the
eigenfunctions with $\vec k =(\pi,0)$ or $\vec k =(0,\pi)$ fulfill the
so-called product-MPSR \cite{Richter:zpb93}. (b) If the number of spins
in the sublattice is odd (e.g. $N=26$), the situation is more
complicated. The energy levels cross each other for $J_2^{c}$ slightly
greater than ${1\over2}J_1$.

\section{Numerical Results}
In subspaces with lower quantum numbers $M < (N/2)-1$ we cannot find
simple expressions for the eigenvalues and eigenfunctions. Hence, we
present exact diagonalization data for $M\le(N/2)-2$. Using a modified
Lanczos procedure we calculate in every subspace $M$ the state with the
lowest energy $E_0(M)$. Since the number of Ising basis states increases
exponentially as ${N \choose N-M}$, one can calculate $E_0(M)$ for {\bf
all} $M={N\over2},\ldots,0$ only for relatively small systems (in our
case $N \le 26$). However, in subspaces with larger $M$ we are able to
present data for $N$ up to $144$. In all cases we use periodic boundary
conditions. Analyzing the eigenfunction with respect to the MPSR we can
determine numerically the critical $J_2^{c}$ where the MPSR is violated.

{\it Linear Chain} - In Fig.\ref{fig1} we show $J_2^{c}$ as a function
of $1/N$. For $M(N) = (N/2)-1$ the analytic result is drawn. For the
next $M(N) = (N/2)-2$ the data show a similar behavior with the same
critical value of $J_2={1\over4}$ for $N\rightarrow \infty$. By
considering the numerical data an analytic solution can be predicted
\be
J_2^c = J_1{1 + \cos \left[ \pi (1- {2 \over (N-1)}) \right] \over
                1 - \cos \left[ 2\pi (1- {2 \over (N-1)}) \right] } .
\ee
The following subspaces $M(N)=(N/2)-p$, $p>2$ show a different behavior
with different critical values for $J_2$ if $N\rightarrow \infty$. These
critical values decrease for increasing $p$ but evidently a finite
region with a non-violated MPSR is preserved.

In Fig.\ref{fig2} the critical  $J^c_2$ is shown as a function of a
renormalized \linebreak $M_r = M(N)/(N/2) $ for small systems
($N=8,\ldots,26$) over the full range of $M_r$. $M_r=1$ is the ground
state subspace for a ferromagnet and $M_r=0$ for an antiferromagnet. The
finite size extrapolation for the ground state with $M_r=0$ yields a
small but finite critical value $J^c_2 \approx 0.03J_1$ which
corresponds to the value estimated by Zeng and Parkinson in
\cite{Zeng:prb95}. The monotonic decrease of $J_2^{c}$ with decreasing
$M_r$ indicates a finite region of a validity of the MPSR for all $M_r$.

{\it Square lattice} - In Fig.\ref{fig3} we show $J_2^{c}$ as a function
of $1/N$. Here the $N$ dependence is less regular and a finite size
extrapolation is much more difficult. This is mainly connected with the
shape of the periodic boundaries. For some of the finite lattices the
boundaries are parallel to the x- and y-axis (e.g. for $N=$4x4,
6x6,...,12x12) whereas for other lattices the boundaries are inclined
(e.g. $N=18,20,32$, see e.g \cite{Oitmaa79}). The impression of an
oscillating behavior of $J_2^c$ versus $1/N$ stems just from the
alternation of parallel and inclined finite lattices. Nevertheless, it
is evident that the critical $J^c_2$ goes to a finite value for
$N\rightarrow \infty$. An extrapolation to the thermodynamic limit for
the antiferromagnetic ground state, i.e. subspace $M=0$, is almost
impossible for the square lattice. However, if we assume for $M=0$ that
the $J_2^c$ is almost independent of $1/N$ for $N>16$ (as it is
suggested by Fig.\ref{fig3} and by spin wave theory) we could estimate
from our data for $N=10,16,18,20$ a critical value of about $0.2 \ldots
0.3$.

Fig.\ref{fig4} supports this estimation. Here the critical $J_2^c$ is
shown for different small lattices $N \le 34$ as a function of $M_r$. It
is seen that for $M_r \le 0.6$ the critical $J_2^c$ does not strongly
depend on $M_r$ (in contrast to the linear chain, where we have a
monotonic decrease) and gives for all the lattices a value of about
$0.3J_1$ for the antiferromagnetic ground state ($M_r=0$).

\section{Conclusion}
For linear chains and square lattices we have shown, that in subspaces
with large quantum number $M$ of the spin operator $S_{total}^z$, the
Marshall-Peierls sign rule is preserved up to a fairly large frustration
parameter $J_2^{c}$.

Moreover, for linear chains the finite size extrapolation is quite
reliable even for the singlet ground state and yields for {\bf all} $M$
a finite parameter region for $J_2$ where the MPSR is valid.

For square lattices we observe in general higher critical values $J_2^c$
compared to linear chains. From this observation and from the
extrapolation for subspaces with higher $M$ we argue that for square
lattices the MPSR is stable against a finite frustration in all
subspaces, too.

\section{Acknowledgments}
This work has been supported by the Deutsche Forschungsgemeinschaft
(Project No. Ri 615/1-2) and the Bulgarian Science Foundation (Grant
F412/94).


\clearpage

\begin{figure}
\caption{Linear chain: The critical value  $J^c_2$ where the MPSR
breaks down versus $1/N$ for different $M$ ($J_1=1$). 
\label{fig1}
}
\end{figure}

\begin{figure}
\caption{Linear chain: The critical value $J^c_2$ where the MPSR
breaks down versus $M_r$ for different $N$ ($J_1=1$).
\label{fig2}
}
\end{figure}

\begin{figure}
\caption{Square Lattice: The critical value $J^c_2$ where the MPSR
breaks down versus $1/N$ for different $M$ ($J_1=1$).
\label{fig3}
}
\end{figure}

\begin{figure}
\caption{Square Lattice: The critical value $J^c_2$ where the MPSR
breaks down versus $M_r$ for different $N$ ($J_1=1$).
\label{fig4}
}
\end{figure}

\end{document}